# The management of scientific and technological infrastructures: the case of the Mexican National Laboratories

L. Munguía[1], J. C. Escalante[2] and E. Robles Belmont[2]

[1]Consejo Nacional de Ciencia y Tecnologia, DADC, DRHCIC, C.P. 03940, Cd. De Mexico, Mexico. Email:leonardoo.munguia@gmail.com

[2]Universidad Nacional Autonoma de Mexico, IIMAS, DMMSS, Apdo. Postal 20-126, 04510 Cd. de Mexico, Mexico.

**Keywords:** Scientific and technological infrastructure, National Laboratories, management, evaluation, metrics, effectiveness

**Abstract**

The effectiveness of research units is assessed on the basis of their performance in relation to scientific, technological and innovation production, the quality of their results and their contribution to the solution of scientific and social problems. This paper examines the management practices employed in some Mexican National Laboratories in order to identify those that could explain their effectiveness in meeting their objectives. The results of other works that propose common elements among laboratories with outstanding performance are used and verified directly in the field. Considering the inherent complexity of each field of knowledge and the socio-spatial characteristics in which the laboratories operate, we report which management practices are relevant for their effectiveness, how they contribute to their consolidation as fundamental scientific and technological infrastructures, how these can be translated into indicators that support the evaluation of their performance, and still pending.

**Introduction**

Mexican National Laboratories are fundamental scientific and technological infrastructures for research and technological development in priority areas for the State. Their distinctive characteristics are that they are financed by the government, have highly specialized research equipment, and provide quality services to the academic, governmental, social and industrial sectors. They were established in 2006 as a result of a call published by the National Council of Science and Technology (Conacyt) with the objective of forming and consolidating National Laboratories, which is generally published annually and has currently supported 90 laboratories.

The incipient development of the laboratories has not allowed us to evaluate their performance. Likewise, the management actions followed within the National Laboratories are not sufficiently known, thus wasting valuable experience that could be used to establish general management guidelines and indicators to evaluate and improve their effectiveness.

In this work, we review the management practices of the laboratories, to answer the question: what are the management practices used in the National Laboratories to ensure the performance of their activities and the fulfillment of their objectives?

For the study of these practices, we used the framework proposed by Jimenez et al. (2018) that proposes four elements that could explain the effectiveness of the National Laboratories (*Laboratory Experience, Network, Work Team Expertise* and *Leadership*). Questionnaires and interviews were applied directly to some members of 10 laboratories and the information collected was examined with Confirmatory Factor Analysis (CFA), Multiple Correspondence Analysis (MCA) and Social Network Analysis (SNA).

We found the following: the *Experience of the laboratories*, although explained by the time of operation and the capacity they have to equip themselves, develop cutting edge and high quality scientific and technological production, is justified by the infrastructure, institutional support and capacity to manage financed projects; the *Network* of the laboratories shows that their relationships with other actors are explained by their thematic nature and spatial proximity, which allows them to have more and better links. In the *expertise of the work team*, although the individual and collective achievements of the laboratory members are identified as central elements, there is also a condition that crosses these results: the link established between the laboratory equipment and the development of the personnel themselves. Finally, the *Leadership* practices are recognized as fundamental, although in the field there is no characteristic type of leadership and there is a combination of ways of leading people.

The structure of this paper is as follows: section 1 characterizes the Mexican National Laboratories and their situation up to 2019; section 2 presents the analytical framework for the study of the practices associated with their effectiveness and presents the laboratories studied; section 3 discusses the methodologies and techniques for analyzing the information, the collection of such information, its processing and the results; section 4 discusses the results, indicates their scope and limitations and proposes some general elements to be considered in the management of laboratories that could be significantly important when developing indicators to evaluate their effectiveness; finally, the conclusions reached by this exploratory exercise are offered.

# 1. National Laboratories as Scientific and Technological Infrastructures

National Laboratories in Mexico are public scientific and technological infrastructures that integrate human, material, financial, technological, and scientific resources from universities, research centers and other higher education and research institutions around strategic research topics defined in the government's agenda such as Environment, Knowledge of the Universe, Sustainable Development, Technological Development, Energy, Health and Society (Conacyt, 2013).

Currently, there are 90 National Laboratories operating which are managed by prestigious working groups in universities, research centers and institutes of the federal and state order in all thematic areas, predominantly in those referring to technological development, health, knowledge of the universe and the environment, as shown in Table 1.

Table 1. National Laboratories by research topic.

| Research topic | Number of National Laboratories |
| --- | :---: |
| Technological development | 37 |
| Health | 15 |
| Knowledge of the universe | 14 |
| Environment | 9 |
| Sustainable development | 6 |
| Energy | 6 |
| Society | 3 |
| **All** | **90** |

Source: Own elaboration based on the National Laboratories Register (Conacyt, 2020).

In addition to these thematic vocations, laboratory sites are concentrated in the Centro-Bajío corridor of Mexico, as shown in Figure 1 below, and some have facilities in more than one state; Michoacán, Estado de México, Morelos, Puebla, Ciudad de México, Querétaro, Guanajuato, San Luis Potosí and Nuevo León; that is, 9 states, concentrate 82.22% of the sites. The rest is distributed among 12 other states and 11 do not have any. The institutions with the most laboratories are the Universidad Nacional Autónoma de México (UNAM), the Centro de Investigación y de Estudios Avanzados del Instituto Politécnico Nacional (Cinvestav), the Instituto Politécnico Nacional (IPN), the Instituto Potosino de Investigación Científica y Tecnológica (IPICYT), and the Universidad Autónoma de Nuevo León (UANL), which concentrate 53 of the 90 laboratories; that is, 58.8% of the total (Conacyt, 2020); the UNAM itself has 33.

Figure 1. Distribution of the National Laboratories' locations along the country.

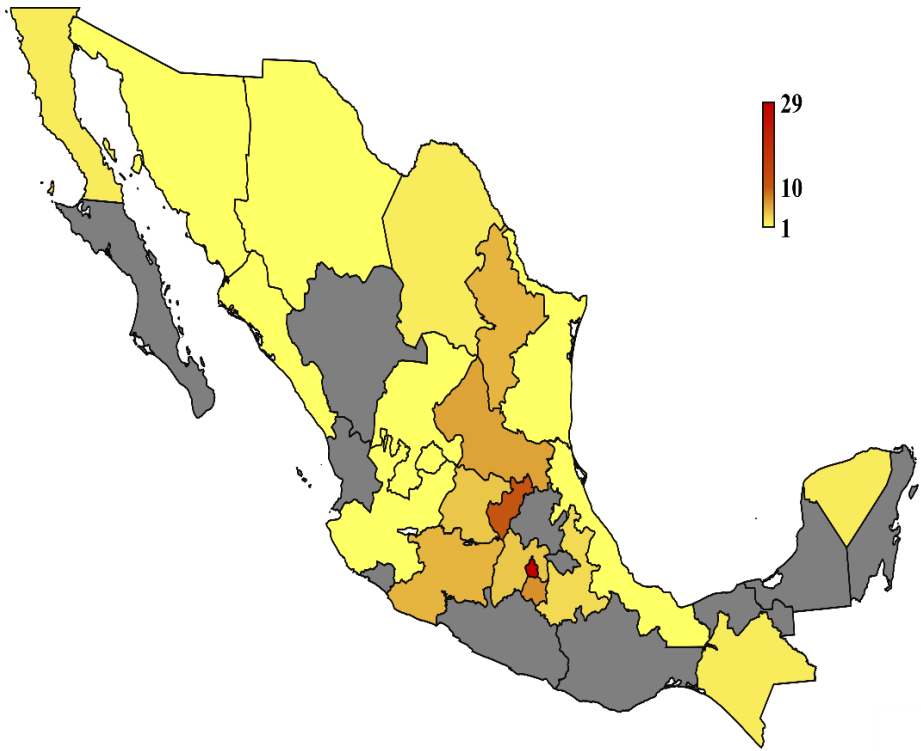

Source: Own elaboration based on the National Laboratories Register (Conacyt, 2020).

The 90 laboratories are totally or partially financed by the Mexican government through the National Council of Science and Technology (Conacyt), which generally since 2006, publishes an annual call for complementary economic support for the acquisition and maintenance of specialized equipment and adaptations of physical spaces, with the specific objective of forming and consolidating National Laboratories which are conceived as "specialized units to reinforce infrastructure and equipment for scientific development and innovation in fundamental areas, in order to optimize resources, generate synergies and offer constant and quality services" (Conacyt, 2020).

On the way to the consolidation of National Laboratories as scientific and technological infrastructures, Conacyt promotes that they perform four functions: research, human resources training, service provision and linkage. These activities increase the social benefits of investment in research and development, expand the scientific and technological capabilities of national Science, Technology and Innovation (STI) institutions in all areas of knowledge and offer an opportunity for laboratories to develop within a framework of economic, social and environmental sustainability.

**1.1 The problem of financing scientific and technological infrastructures**

After 15 years of governmental support, today laboratories are in different stages of development. All of them do or support research, human resources training, service provision and serve as strategic platforms for innovation in the academic, industrial, social, and governmental sectors at a national and international level. Moreover, laboratories require continuous financial resources to develop their research and scientific and technological development projects which are generally used to acquire new equipment, ensure its operation, maintenance and updating.

The National Laboratories mainly resort to public funds from Conacyt to cover their demand for financial resources, as mentioned above, and to a lesser extent to allocations from the universities, research centers and institutes where they are located. This situation of generalized dependence of the institutions, particularly the laboratories, on government spending in the current context of investment in the Mexican Science and technology sector, is characterized by a persistent and historical scarcity of economic resources[1], compromising

---

[1] As evidenced by the indicator of Spending on Scientific Research and Experimental Development (SSRED) which during the period 2010-2019 represented less than 0.5% as a proportion of Mexican Gross Domestic

the development of the National Laboratories since it affects the availability of the financial resources that are necessary mainly to maintain and strengthen them.

For the above reasons, this paper is interested in knowing the management actions of the National Laboratories used by their working groups, for which we will seek to answer the question: what are the management practices used in the National Laboratories to ensure the realization of their activities and the fulfillment of their objectives?

This paper employs the framework proposed by Jiménez et al. (2018) to explain the "success" of the laboratories; that is, their effectiveness in carrying out their activities and meeting their objectives for which the actions associated with this success in 10 of the laboratories are reviewed. The framework and the methodological strategy followed for its operation are shown below.

## 2. Reference framework and empirical strategy

It is recognized that each National Laboratory has its own scientific and technological culture (Minhot and Torrano, 2009). This dissimilarity makes it difficult to identify operational schemes that prove to be effective for all. However, as scientific and technological infrastructures, they respond to local-global pressures (Prahalad and Doz, 1987) and face common challenges such as their strengthening and advancing towards the future with self-sufficiency.

---

Product (GDP) and experienced a reduction to its funding by the government and business sectors in real terms of 14.26% and 60.69%, respectively (Conacyt, 2019).

In order to understand how they articulate their actions to respond to the demands of their environment with a long-term vision, the practices developed in the laboratories that are associated with their effectiveness in meeting their objectives are studied. In this sense, the reference framework proposed by Jiménez et. al (2018) is useful, as it is focused on the work of National Laboratories. The authors indicate that they have identified 4 elements common to successful laboratories:

1. *Laboratory experience*. Laboratories with extensive experience are more likely to be successful in their new role as National Laboratories.

2. *Network*. Laboratories with an extensive network have an advantage in reinforcing already established connections as opposed to laboratories that start from scratch in building a strong network.

3. *Team expertise*. Those laboratories formed with expert teams conduct innovative research.

4. *Leadership*. The leader must be an internationally recognized scientist and must have not only the acceptance of the group but also the desire and courage to work together.

The *laboratory experience* refers to the capacity of laboratories to mobilize nature and produce scientific facts which implies having certain material and institutional conditions that are the result of a "series of institutional modifications aimed at the material, social and symbolic conformation of a space of scientific production concretized in laboratories" (Arellano and Ortega, 2002). The working groups that have set up laboratories in universities, research centers and institutes should at least have institutional support and commitment, as well as the capacity to obtain the material and economic resources to equip, maintain and operate them on an ongoing basis, usually on the basis of funded projects.

Laboratories form networks of actors that depend on the principle of preferential attachment. "Preferential attachment" is a rule to become richer with a network's growth. A node with many existing connections is more likely to gain new connections" (Honner, P., 2018). In this sense, the existence of a collaborative *Network* is critical for its development since the effectiveness of a laboratory depends not only on the management and efficiency in each of its phases but also on linking with key actors during the processes of scientific and technological development. These connections contribute to raise its recognition and competitiveness.

Laboratories are composed of *researchers, technicians, and students with expertise* who "are committed to the line of research, patterns of activity organization, certified production of the group and the fields in which scientific activity is deployed" (Arellano and Ortega, 2002), and who share "beliefs, habits, systematized knowledge, exemplary achievements, experimental practices, oral traditions and craft skills" (Latour and Woolgar, 1986).

Empirical evidence shows that "the leader in research units has a considerable influence on the planning of research activities and on the integration between research strategies and structures, and these two roles have a major influence on the climate of the organization, which has a direct effect on research effectiveness" (Knorr et. al., 1979). In other words, the "leader is one of the most important influencing factors in the performance of research units because of his contribution to the integration, atmosphere, quality of the research program and its external links" (Nagpaul, S., 1990). For this reason, laboratories require working groups and *leaders* that allow them to produce knowledge and technologies, reproduce human capital and provide quality services that allow them to obtain resources to advance towards financial, social, and environmental self-sufficiency.

According to the above, laboratories that have a priori a certain degree of development in terms of infrastructure, project management capacity, institutional support, some collaboration networks with key actors, availability of highly specialized human resources and leaderships that have proven a positive performance in relation to scientific and technological production would have greater possibilities of contributing with knowledge and technologies to solve the problems of science and those of national-state priority within a framework of economic, social and environmental sustainability.

To investigate these elements, we examined some of the practices associated with the laboratories through the application of questionnaires and interviews with the researchers in charge of them (scientist responsible/leader) and their collaborators. These instruments were designed based on a review of the literature on the aspects that are of interest in this work so that their results also offer validation of the theoretical construct from which the National Laboratories are being studied.

The information collected with the questionnaires was processed with the Confirmatory Factor Analysis (CFA), Multiple Correspondence Analysis (MCA), Social Network Analysis (SNA) and complemented with the content analysis of the interviews applied. 10 laboratories administered by the Universidad Nacional Autónoma de México (UNAM) were studied, these are:

I. Laboratorio Nacional de Clima Espacial (National Laboratory for Space Climate, LANCE). Topic: Environment.

II. Laboratorio Nacional de Innovación Ecotecnológica para la Sustentabilidad (National Laboratory for Eco-technological Innovation in Sustainability, LANIES). Topic: Technological development.

III. Laboratorio Nacional de Manufactura Aditiva, Digitalización 3D y Tomografía Computarizada (National Laboratory for Additive Manufacture, 3D Digitalization and Computerized Tomography, MADIT). Topic: Technological development.

IV. Laboratorio Nacional de Resonancia Magnética e Imagenología (National Laboratory for Magnetic Resonance and Imaging, LANIREM). Topic: Health.

V. Laboratorio Nacional HAWC de Rayos Gamma (HAWC Gamma Rays National Laboratory). Topic: Knowledge of the Universe.

VI. Laboratorio Nacional de Materiales Orales (National Laboratory for Oral Material, LANMO). Topic: Society.

VII. Laboratorio Nacional de Ciencias de la Sostenibilidad (National Laboratory for Sustainability Sciences, LANCIS). Topic: Sustainable development.

VIII. Laboratorio Nacional de Ciencias para la Investigación y Conservación del Patrimonio Cultural (National Laboratory for Sciences in Research and Conservation of Cultural Heritage, LANCIC). Topic: Society.

IX. Laboratorio Nacional en Salud: Diagnóstico Molecular y Efecto Ambiental en Enfermedades Crónico-Degenerativas (National Laboratory for Health:

    Molecular Diagnostics and Environmental Effect on Chronic-Degenerative Illness, LNS-FESI). Topic: Health.
X. Laboratorio Nacional de Visualización Científica Avanzada (National Laboratory for Advanced Scientific Visualization, LAVIS). Topic: Technological development.

## 3. Information, analysis, and results

The study in the National Laboratories was conducted in two stages. The first collected the information concerning the categories of *Laboratory Experience* and *Network* in 2018. The second stage consisted of collecting what was needed for the categories of *Work Team Expertise* and *Leadership* in 2019. In both cases, field visits were conducted and the two aforementioned instruments were applied to the working groups.

  For the categories *Laboratory Experience* and *Network*, the instrument was applied to the Scientist Responsible. For the categories *Expertise of the work team* and *Leadership*, the instrument was applied to the Scientist Responsible and specifically to a couple of their laboratory collaborators. The information collected from the interviews regarding *Laboratory and Work Team Experience* was examined with content analysis. The information on the *Network* was analyzed through the SNA and that regarding *Leadership* with the CFA and the MCA.

  In the case of *Laboratory Experience*, the analysis of the interviews confirms that those laboratories that have the capacity to manage funded projects, institutional support, and equipment prior to their insertion in the Conacyt program have greater opportunities to perform better in their new role as National Laboratories. Many of the laboratories that were awarded the distinctions already had a previous trajectory between 3 and 10 years. They have, therefore, certain technical conditions, links, and recognition. The HAWC national laboratory, for example, is a laboratory that already had technical and instrumental capabilities, institutional recognition, and international prestige prior to its distinction as a National Laboratory.

Likewise, this experience is also fundamental at the individual and collective level of the laboratory members, since in the experience of the work team, the more social capital (Bourdieu, 1987) the individual members of the laboratory have, the greater the probability of effectiveness in carrying out the scientific work. Here it was also identified that the training of laboratory collaborators is strengthened and expanded with the availability of specialized equipment, so that the latter also plays an important role in the training and the pool of highly qualified personnel.

In the case of the network, the SNA[2] used the degree centrality measure to show the importance of the actors structurally as well as a community analysis using Louvain's method to extract groups from the modularity of the edges in the network[3]. Then, the Kamada-Kawai (1988) algorithm was applied for the visualization of the analysis[4].

---

[2] The SNA studies reticular information from a set of actors and provides local and structural numerical measures and visual representations.

This SNA of the collaborations of the 10 National Laboratories studied shows that LANCIS and LANCIC have the largest networks, as illustrated in Figure 2. In general, the laboratories are related to entities from all sectors (academic, industrial, governmental and social) although the academic and governmental sectors are the most referenced. The structure of the network shows that the collaborations of the laboratories have been established with other actors of similar nature with respect to their thematic vocation, although they have also been established with spatially close actors.

The laboratories LNS-FESI, LANIREM and LAVIS form a community and collaborate mainly with other health sector actors of a national governmental and academic nature; LANCIS and LANCE also collaborate with national governmental and academic actors associated with their areas of incidence; LANMO, LANCIC and LANIES articulate with actors from the academic and social sector also from the national sphere, and MADIT and HAWC with actors from the academic sector and mainly from the international sphere. The weakness of the links of these laboratories with actors in the industrial or business sector is a constant feature.

---

[3] They were assigned colors indicating communities of actors with the maximum modularity in their relationships or the greatest structural cohesion.

[4] This algorithm works a model of springs in a dynamic system that relates "the balance of the vertex trace to the balance of the springs of a dynamic system. As a result, the degree of imbalance can be formulated as the total energy of the springs" so that the diagrams that best show the structure of the relationships are obtained by decreasing the energy. The best representation is the minimum energy state of the system. In this model, the original length of the springs (edges) corresponds to the desirable length between the vertices in the diagram and the distance between the vertices of the network is defined as the length of the shortest paths between them. Here "the density of the particles [vertices] is not large because each pair of nodes is forced to maintain a certain distance by the tension [energy] of the spring". In this sense, the illustration shows the best diagram keeping the theoretical properties of the graph in the plane. The result can be interpreted visually, where the size and location of the vertices (actors) indicate their level of centrality and importance in terms of theoretical-graphical closeness to each other.

Figure 2. Network of collaborations of the National Laboratories studied.

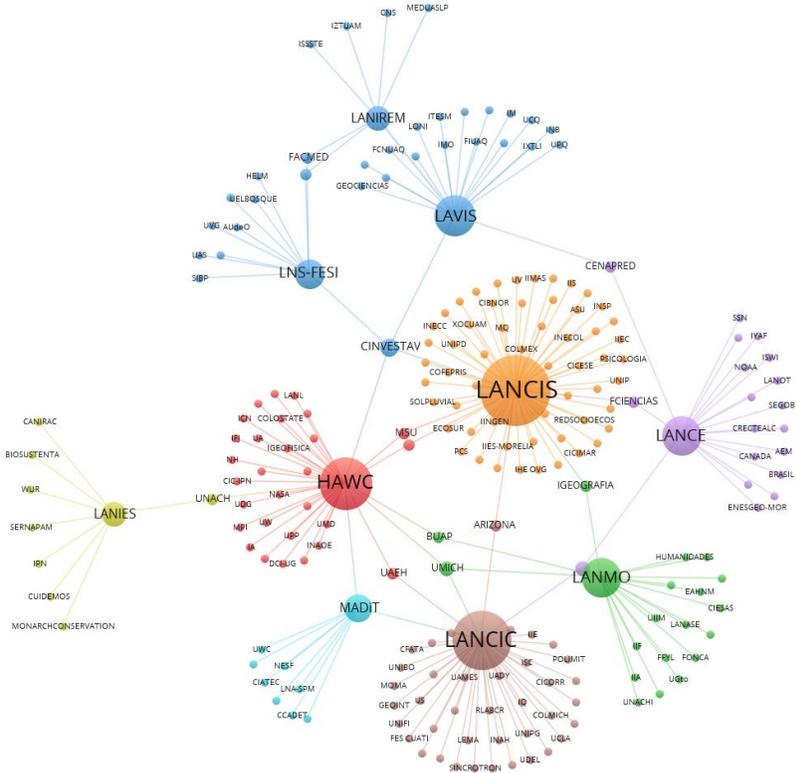

Source: Data obtained from the questionnaire applied to the National Laboratories in 2018.

Regarding the study of *Leadership*, the practices through which the Scientist Responsible (or leader) of the laboratories influence the performance of activities concerning scientific work according to the particular objectives of these scientific and technological infrastructures were investigated. A typology for the study of leadership in research units was used for developing countries, particularly for the Mexican case, which considers *authoritarian, paternalistic, technocratic,* and *democratic* styles (Jiménez, 1990).

The practices of each style were characterized and transferred to a questionnaire with responses on a Likert scale 1-5 to assess frequencies and measure of agreement/disagreement. The questionnaire was applied, and its validation was carried out with the CFA[5] for the validation of constructs regarding the measures of it. The indicators used were the Comparative Fit Index (CFI) and the Tucker Lewis Index (TLI)[6] and of reliability Cronbach's (Nunnualy, J., & Bernstein, I., 1994) and McDonald's Ω (1999). The variables[7] that would best explain each type of leadership were found, namely:

---

[5] The CFA "is a type of Structural Equation Modeling (SEM) that provides measures of the relationships between indicators and latent variables or factors" (Brown, T., 2006). This analysis requires the sample covariance matrix between the manifest variables (observed variables) as input, which identify the latent variables (or leadership styles). By means of the CFA, the latent variables are constructed from the manifest variables incorporating the covariance structure. The constructs that are generated identify the factor loadings that are used to generate the scores of the latent variables (Bollen, K., 1989; Brown, T., 2006).

[6] These "compare the fit of a target model with the fit of an independent or null model (compare the proposed model against the null model) and indicate that the model of interest improves the fit by 95% relative to the null model" (Kline, R., 2005).

[7] The coefficients associated with each manifest variable are called "factor loadings". Factor loadings indicate the loading of each variable on each factor, and these describe the structural relationship between a latent variable and an observed variable, typically reflecting the correlation structure from the covariance matrix. This analysis also provides an error measure, or residual, of the observed variables with the latent variables. The general CFA model is as follows:

$$x = \Lambda x\ \xi + \delta,$$

Table 2. Standardized factor loads of the authoritarian style variables.

| Variables | The Scientist Responsible... Description | Loads |
|---|---|---|
| q6 | Monitors group discussions | 0.57 |
| q9 | Provides feedback after decisions have been made | 0.55 |
| q17 | Personally directs the completion of lab activities | 0.81 |
| q23 | Requires activities to be done as specified by him | 0.68 |
| q32 | Requests unquestioning obedience | 0.60 |

α= 0.77; Ω=0.78; CFI = 1; TLI= 1

Source: Data obtained from the applied questionnaire to the National Laboratories (2019).

Table 3. Standardized factor loads of the paternalistic style variables.

| Variables | The Scientist Responsible... Description | Loads |
|---|---|---|
| q2 | Takes into account the interests of employees | 0.58 |
| q5 | Explains decisions once they have been made | 0.71 |
| q11 | Delegates functions | 0.62 |
| q16 | Provides support, protection, and care | 0.75 |
| q31 | Is interested in the well-being of the employees | 0.90 |

---

where x are the observed variables, ξ are the latent variables or dimensions, δ are the measurement errors and Λx is the vector containing the factor loadings of the latent variables on the manifest variables. Additionally, some reliability and fit parameters emanate from this analysis, where the most used in reliability are Cronbach's α (Nunnualy, J., & Bernstein, I., 1994) and McDonald's Ω (1999). These reliability measures assume that the indicators used are of the concept when they meet the condition of being greater than 0.7 (Tavakol, M., & Dennik, R., 2001). The α-Cronbach measures the internal consistency of the measure and takes values between 0 and 1 and expresses the sum of covariances between the components of a linear combination whose ratio estimates the variance of the sum of all the elements of the variance and covariance matrix (Ibidem, 2006). The coefficient Ω or composite reliability coefficient is estimated from the factor loadings and its value is directly proportional to the value of the loadings.

$\alpha= 0.83; \Omega=0.84; CFI = 0.93; TLI= 0.85$

Source: Data obtained from the applied questionnaire to the National Laboratories (2019).

Table 4. Standardized factor loads of the technocratic style variables.

| Variables | The Scientist Responsible... Description | Loads |
|---|---|---|
| q3 | Listens to feedback and incorporates it into technical decisions | 0.84 |
| q7 | Consults on technical issues | 0.79 |
| q28 | Assesses efficiency in laboratory operation | 0.94 |

$\alpha= 0.87; \Omega=0.89; CFI = 1; TLI= 1$

Source: Data obtained from the applied questionnaire to the National Laboratories (2019).

Table 5. Standardized factor loads for democratic style variables.

| Variable | The Scientist Responsible... Description | Loads |
|---|---|---|
| q4 | When making decisions, he calls upon the collaborators | 0.62 |
| q10 | Encourages participation and free discussion | 0.88 |
| q12 | Gives guidance when in doubt | 0.76 |
| q20 | Reminds the group of collective responsibilities | 0.66 |
| p22 | Distributes responsibility | 0.74 |

$\alpha= 0.83; \Omega=0.85; CFI = 0.88; TLI= 0.77$

Source: Data obtained from the applied questionnaire to the National Laboratories (2019).

These variables were taken, the score was calculated for each of the laboratories and the MCA[8] was performed. The point cloud with the results of this analysis is shown in Figure

---

[8] Multiple Correspondence Analysis (MCA) "is a method of data analysis that graphically represents tables of data" (Greenacre, M., 1993) and allows us to obtain interpretations that show the association between the rows and columns of a contingency table; that is, a table with qualitative variables (of the nominal or ordinal type). The representation of the data in contingency tables is seen as a cloud of points in two dimensions whose relative positions are established on the basis of the value of each of the variables, with each position reflecting the

3. The leadership styles identified within the laboratories are not pure, since the practices observed in personnel management have characteristics of more than one type.

The LANIREM, HAWC and LANCIS laboratories show practices that are closer to the technocratic and paternalistic styles while LANCE is close to democratic leadership and MADIT corresponds to the authoritarian style. On the other hand, the styles exercised in LANMO, LANIES, LAVIS, LN-FESI and LANCIC are not defined under this metric.

Figure 3. Leadership styles exercised in the National Laboratories studied.

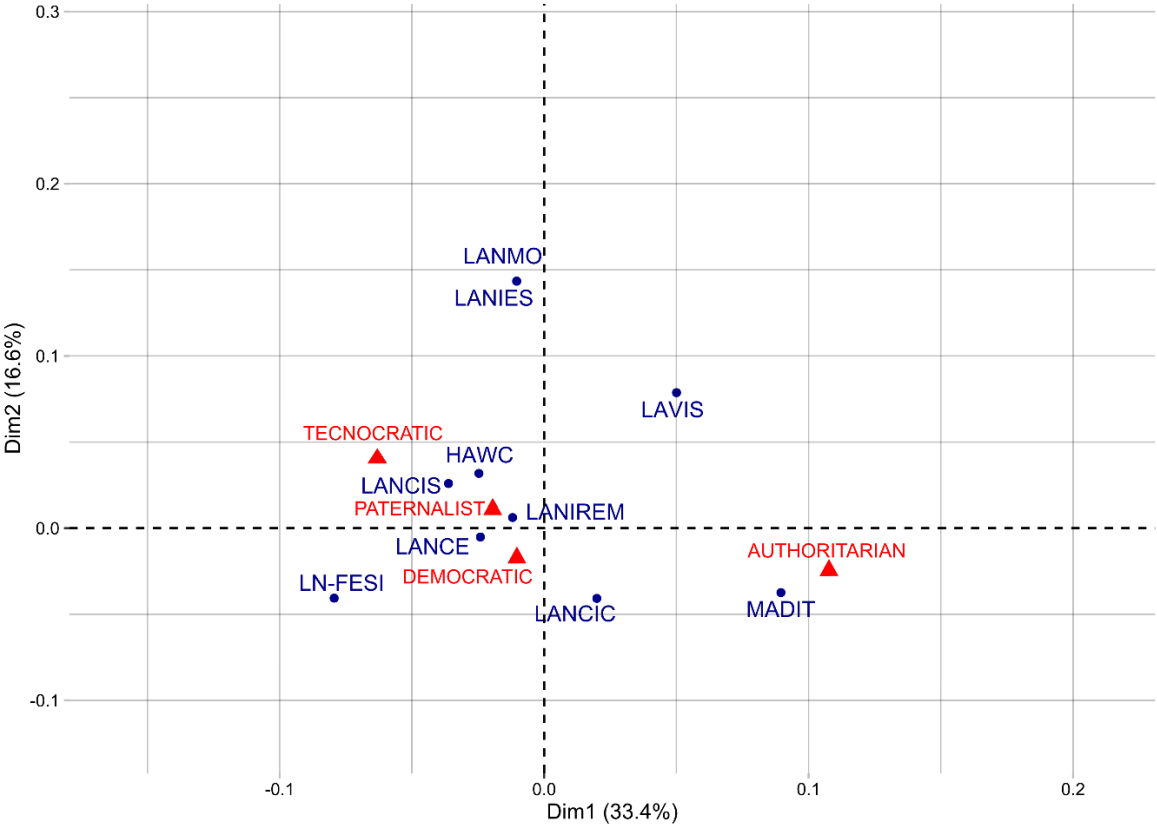

Source: Data obtained from the applied questionnaire to the National Laboratories (2019).

degree of association between each variable. This means that proximity in the graphical representation means correspondence between categories.

## 4. Discussion

The related practices that explain the threshold of the *Laboratory Experience* that were identified exhibit the management capacity of funded projects by the Scientist Responsible and the facilities provided by the institutions to carry out these projects, primarily the acquisition, maintenance, and operation of specialized equipment, in which institutional support and commitment play a central role. Decision-makers and institutions with National Laboratories could consider establishing work schemes to strengthen the management capacity of funded projects and the critical routes to reinforce their technical-scientific capabilities.

*The expertise of the work team* is explained by practices that promote the expansion of the pool of knowledge and individual skills of laboratory members. This is directly related to the availability of sufficient, modern, and adequate equipment for scientific work as well as the recruitment of new talent. Scientific equipment is thus intertwined with capacity building of staff and attraction of other members. The practices associated with this element denote the level of qualification of the members, their results in the laboratory, both individually and collectively, and their contribution to the training of new scientists and technologists, as do the links with other entities to access other scientific and technological infrastructures and exchange experience and personnel. Therefore, we consider that the evaluation of the *expertise of the working team* should also consider the accessibility to other equipment and the existence of collaboration instruments with other institutions.

Continuing with the *Network* factor, the actions tending to exploit relationships favored by conditions such as the thematic nature and spatial proximity are remarkable.

These patterns suggest some regularities regarding the vocations of the laboratories and the types of entities with which they relate, as shown by the most frequently observed proximities between laboratories and academic and governmental entities. In this line, the laboratories that have more key collaborations in services, technology transfer and scientific and technological production are indirectly linked to other actors with which establishing relationships could increase their recognition and competitiveness. Thus, the assessment for this dimension could consider the type of entities with which relationships have been established, their vocation, diversity, intensity, means and modes through which a link is developed.

Finally, the personal styles of personnel management used by the Scientist Responsible or leader observed in these scientific and technological infrastructures are varied, although technocratic, democratic, and paternalistic style practices are more frequent. It is considered that the framework used to analyze *Leadership* is not sufficient to characterize the way of leading the personnel of a research unit. It is also useful to characterize the ways in which laboratory personnel organize themselves to carry out scientific work on an individual basis. Observations are needed to broaden the scope of this measurement and another conceptual framework to incorporate other types of horizontal, collaborative, and relational leadership.

In sum, the factors or dimensions that explain the effectiveness or success of the laboratories are neither unique nor definitive, nor are the practices into which they are translated. We insist that this exercise is exploratory and could underestimate (or overestimate) the variables analyzed, especially because there is the influence of variables not sufficiently known that affect the performance of the National Laboratories such as those inherent to the architecture of the national STI institutions and the dynamics intrinsic to the laboratories in their areas of knowledge.

An alternative to apprehend with consistency the practices that promote the effectiveness of the National Laboratories could be the extension of the observations because this would allow to deepen the knowledge about the management strategies of the scientific and technological infrastructures beyond the UNAM, the Mexican Centro-Bajío corridor and the vocations seen in this work.

**Conclusions**

The National Laboratories concentrate resources and efforts of the State to meet the demand for knowledge and technologies and to take advantage of strategic opportunities while increasing the social benefits of investment in STI. After almost 15 years since this important initiative was launched by Conacyt, and in the current context of generalized crisis, it is essential to carry out a diagnosis that allows us to know the capacities and practices of the laboratories that have proven to be effective in the fulfillment of their activities and objectives. This work focused on studying them in a general way because their understanding and dissemination would support decision making with the purpose of strengthening the scientific and technological infrastructure, optimizing resources, generating synergies, and offering quality services within a framework of financial, social and environmental sustainability. In this sense, the present study responds in a fundamental way to the question "what are the management practices used in National Laboratories to ensure the performance of their activities and the fulfillment of their objectives?"

The answer to this question was found in the absence of information regarding to equally successful strategies or practices relating specifically to scientific and technological infrastructures such as National Laboratories. This implied mobilizing the framework proposed by Jiménez et al. (2018) to address the issue directly through the application of questionnaires and interviews to a sample of 10 laboratories based at UNAM. The data collected through these instruments refer to *Laboratory Experience, Network, Work Team Expertise* and *Leadership*. These were analyzed, among others, making use of the CFA, SNA and the MCA.

As a result, the *Laboratory Experience* shows that the National Laboratories have a life span of between 3 and 10 years, even before being labeled as such, during which they have assembled specialized teams and scientific equipment that distinguish them in their regions, in the country or internationally, while also developing a level of cutting-edge scientific and technological production to contribute strategically in the scientific, technological, economic, political, and social fields. In terms of the *Network*, the laboratories most frequently exploit relationships that, by their nature, allow them to take advantage of the conditions of their immediate environment, in the physical, institutional, thematic, and strategic spheres. Regarding the *Expertise of the work team*, it was found that the laboratories are distinguished by the quality of their personnel, as well as the quantity and quality of the laboratory equipment they have, which allows them to perform better in their work. Finally, *Leadership* in the laboratories is characterized by technocratic, democratic, and paternalistic practices in decision making.